\documentclass{article}
\usepackage{frascatiphys,here,graphicx,subfigure}
\newcommand{\la}{\langle}
\newcommand{\ra}{\rangle}

\newcommand{\beq}{\begin{eqnarray}}
\newcommand{\eeq}{\end{eqnarray}}

\newcommand{\PLB}{Phys.\ Lett.\ {\bf B}}
\newcommand{\PTP}{Prog.\ Theor.\ Phys.~}

\newcommand{\PRD}{Phys.\ Rev.\ {\bf D}}
\begin{document}
\title{
THE \mbox{\boldmath $q\bar{q}$} 
\mbox{\boldmath $S$}-WAVE AXIAL-VECTOR MESONS \\
IN THE COVARIANT \mbox{\boldmath $\widetilde{U}(12)$}-SCHEME
}
\author{
Tomohito Maeda, Kenji Yamada, \\
{\em Department of Engineering Science, 
Junior College Funabashi Campus, } \\
{\em Nihon University, Funabashi 274-8501, Japan} \\
Masuho Oda,\\
{\em School of Science and Technology,} \\
{\em Kokushikan University, Tokyo 154-8515, Japan}\\
Shin Ishida,~{\thanks{ \ \ Senior Research Fellow}}\\
{\em Research Institute of Science and Technology,}\\
{\em College of Science and Technology,} \\
{\em Nihon University, Tokyo 101-8308, Japan}
}
\maketitle
\baselineskip=11.6pt
\begin{abstract}
We study the properties of 
axial vector mesons $a_{1}$ and $b_{1}$ as 
relativistic S-wave states 
which are predicted in 
the $\widetilde{U}(12)$-scheme, 
through the analyses of their radiative and 
pionic decays. 
Specifically, partial widths of the strong 
$a_{1}(b_{1})$ $\to \rho(\omega)\pi$ processes, 
their $D/S$-wave amplitude ratios, and 
radiative transition widths of 
$a_{1}(b_{1}) \to \pi \gamma$ processes are calculated 
by using a simple decay interaction model, and 
made a comparison with the 
respective experimental values. 
\end{abstract}
%%%%%%%%%%%%%%%%%%%%%%%%%%%%%%%%%%%%%%%%
\baselineskip=14pt
%%%%%%%%%%%%%%%%%%%%%%%%%%%%%%%%%%%%%%%%
\section{Introduction}

In recent years we have proposed the 
$\widetilde{U}(12)$-scheme\cite{Cov,u-12}, 
a relativistically covariant level-classification 
scheme of hadrons. 
In this scheme, the ground state (GS) of 
light $q\bar{q}$ meson system is assigned 
as ${\bf 12} \times {\bf {12}^{*}}={\bf 144}$-
representation of the $U(12)_{SF}$-group 
at their rest frame. 
The ${U}(12)_{SF}$-group includes, 
in addition to the conventional non-relativistic 
$SU(6)_{SF}$-group, the new symmetry $SU(2)_{\rho}$
{\footnote{The new degree of 
freedom corresponding to the 
$SU(2)_{\rho}$-symmetry is called 
the $\rho$-spin, after the well-known 
$\rho\otimes\sigma$-decomposition of 
Dirac matrices. }}, 
which corresponds to the degree of freedom associated 
with negative energy Dirac spinor solutions 
of confined quarks inside hadrons. 
By inclusion of this extra $SU(2)$ spin freedom, 
it leads to the possible existence of some extra 
multiples, called 
{\it chiral states}, 
which do not exist in the ordinary 
non-relativistic quark model (NRQM). 
As an example, the light 
scalar $f_{0}(600)$/$\sigma$ meson, 
a controversial particle for long time, 
is identified as 
$S$-wave chiral state 
as well as $\pi$ meson, 
and they play mutually the role of chiral partners 
in the $\widetilde{U}(12)$-scheme. 
As is well known, in conventional level classification 
scheme based on NRQM, 
lowest scalar meson is obliged to be 
assigned as orbital $P$-wave excited state. 
%On the other hand, the $\sigma$ meson 
%as a chiral state 
%originated from 
%the covariant nature of the $\widetilde{U}(12)$-scheme, 
%is as sort of a {\it relativistic $S$-wave state}. 
As an another example, 
the $a_{1}$ meson, possibly 
to be identified as 
the $q\bar{q}$ $S$-wave axial-vector mesons 
in the $\widetilde{U}(12)$-scheme. 
They form a linear representation of chiral symmetry 
with the $S$-wave $\rho$ meson. 
Here it is notable that these $\sigma$ and $a_{1}$ 
mesons are expected to have the light mass compared with 
the conventional case of the $P$-wave states. 
Furthermore, ${\bf 144}$-representation includes 
the another axial-vector 
meson state with $J^{PC}=1^{+-}$, 
to be identified with the $b_{1}$ meson. 
\\
In this work, we try to elucidate the properties of 
our new-type $S$-wave axial-vector 
mesons, $a_{1}$ and $b_{1}$, 
whose existence are predicted in 
the $\widetilde{U}(12)$-scheme, 
through the analyses of their radiative and 
pionic decay. 
In the actual analyses, we identify 
our chiral $S$-wave $a_{1}$ and $b_{1}$ mesons as 
the experimentally well-known states, $a_{1}(1260)$ and 
$b_{1}(1235)$, respectively. 
Then, by using a simple decay interaction, their 
partial widths of the strong 
$a_{1}$ ($b_{1}$)$\to \rho$ ($\omega$)$\pi$ decays 
(with $D/S$-wave amplitude ratios, ) and 
radiative transition widths of 
$a_{1}$ ($b_{1}$) $\to \pi \gamma$ 
processes are calculated 
in comparison with 
the respective experimental values. 
%%%%%%%%%%%%%%%%%%%%%%%%%%%%%%%%%%%%%%%%%
\section{Wave functions of the \mbox{\boldmath $a_{1}$} 
and \mbox{\boldmath $b_{1}$ mesons 
as $S$-wave chiral states} }
In this section we collect the concrete expressions 
of meson wave function 
(WF) in our scheme 
necessary for the relevant applications
%and recapitulate their physical meaning briefly
\footnote{In more detail, 
see Ref. \cite{Cov,u-12,Dshep}}. \\
The basic framework of our level-classification scheme is 
what is called the boosted LS-coupling (bLS) scheme. 
In this scheme, the WF of 
$q\bar{q}$ GS mesons are 
given by the following (bi-local Klein Gordon) 
field with one each upper and lower indices
{\footnote{
For simplicity, the only positive frequency part of WF 
is shown here.
}}, 
\beq
\Phi(X,x)^{(+)}_{A}{}^{B} = N e^{+iP \cdot X} \ 
W(v)^{(+)}_{\alpha, a}{}^{\beta, b}
\ f_{G}(v,x).
\label{Eq1}
\eeq
Where $A=(\alpha,a)$ ($B=(\beta,b)$) denotes 
Dirac spinor and flavor indices
respectively, $X_{\mu}$ ($x_{\mu}$) represents 
the center of mass (CM) (relative) coordinate 
of the composite meson. The ${P_{\mu}}$ 
($v_{\mu}=P_{\mu}/M$, $M$ being the mass of meson; 
$v_{\mu}^2=-1$, $v_{0}=+1$) 
denotes 4-momentum (4-velicity) of the relevant mesons. 
In the bLS scheme, respective 
spin ($W(v)_{A}{}^{B}$) and 
space-time
{\footnote{
We have been adopted a 
definite metric type 4-dimensional oscillator 
function as $f_{G}(v,x)${\cite{MassCOQM}}.
}} 
($f_{G}(v,x)$) parts of WF 
are, separately, 
made covariant by boosting 
from the corresponding parts of NR ones. \\
Important feature of 
$\widetilde{U}(12)$-scheme is that 
the spin WF contains 
extra $SU(2)$ spin degree of freedom, 
called $\rho$-spin.
As expansion bases of spinor WF, we use 
the Dirac spinor with hadron on-shell 4-velocity, 
\beq
\{ u_{+}(v), u_{-}(v) \}. \ \ \ \ 
(\rho_{3} \ u_{\pm}=\pm u_{\pm})
\eeq
Here, $u_{+}$ corresponds 
to conventional constituent 
quark degree of freedom, while $u_{-}$ is 
indispensable for 
covariant description of confined quarks
{\footnote{They form the chiral partner 
in basic representation of the chiral group.} }. 
%%\footnote{
%$\gamma_{4} u_{\pm}(v) = \pm u_{\pm}(v)$, 
%$\gamma_{5} u_{+}(v) = - u_{-}(v)$
%}
Accordingly, expansion basis of $q\bar{q}$ meson WF 
is given by direct product 
of the respective spinor WF corresponding to 
the relevant constituent quark. 
They consist of totally 16 members 
in $\tilde{U}(4)_{S}$-space as, 
\beq
W(v)_{\alpha}{}^{\beta} = 
u_{r}(v)_{\alpha}\bar{v}_{r'}(v)^{\beta}. \ \ \ \ 
(r,r^{'})=(\rho_{3},\bar{\rho}_{3})
\eeq
We show the specific form of spin WF for 
the respective members of $q\bar{q}$ 
$S$-wave mesons, appeared in the relevant 
applications, in Table {\ref{tab1}}. 
Here it should be noted that, 
in the actual application, 
being based on its 
success{\cite{Oda}} with $SU(6)_{SF}$-description 
for $\rho(770)$-nonet, 
it seems that its WF 
should be taken as the form 
containing only positive $\rho_{3}$-states. 
This is made by taking the 
equal-weight superposition of 
two spin WF which belongs to the different 
chiral representation, respectively. 
%%%%%%%%%%%%%%%%%%%%%%%%%%%%%%%%%%%%%%%
\begin{table}[htbp]
\centering
\caption{ {\it Spin wave functions of $S$-wave mesons 
applying for in this work, at their rest frame. Note that 
the physical $\rho$ meson WF is given by as a sum of the 
following two vector WF, 
the one only with $(\rho_{3},\bar{\rho}_{3})=(+,+)$. 
}~(i=1,2,3)
}
\vskip 0.1 in
\begin{tabular}{|l|c|c|c|c|} \hline
Mesons & $J^{PC}$ & $W(v=0)^{(+)}$& $SU(2)_{L}\otimes SU(2)_{R}$ &$(\rho_{3},\bar{\rho}_{3})$ \\
\hline
\hline
$a_{1}(1260)$& $1^{++}$ & $\frac{\gamma_{5}\gamma_{i}}{2}$&$(1_{L},0_{R})\oplus (0_{L},1_{R})$&$\frac{(-,+)+(+,-)}{\sqrt{2}}$\\ \cline{1-5}
$b_{1}(1235)$& $1^{+-}$ & $\frac{i\gamma_{5}\sigma_{i4}}{2}$&$({\frac{1}{2}}_{L},{\frac{1}{2}}_{R})$&$\frac{i\left((-,+)+(+,-)\right)}{\sqrt{2}}$\\ \cline{1-5}
$\rho(770)$& $1^{--}$ & $\frac{i\gamma_{i}}{2}$&$(1_{L},0_{R})\oplus (0_{L},1_{R})$&$\frac{(+,+)+(-,-)}{\sqrt{2}}$\\ \cline{2-5}
$\rho(1250)${\cite{Yamauchi}}& $1^{--}$ & $\frac{\sigma_{i4}}{2}$&$({\frac{1}{2}}_{L},{\frac{1}{2}}_{R})$&$\frac{(+,+)-(-,-)}{\sqrt{2}}$\\ \cline{1-5}
$\pi(140)$& $0^{-+}$ & $\frac{i\gamma_{5}}{2}$&$({\frac{1}{2}}_{L},{\frac{1}{2}}_{R})$&$\frac{(+,+)+(-,-)}{\sqrt{2}}$\\
\hline
\end{tabular}
\label{tab1}
\end{table}
%%%%%%%%%%%%%%%%%%%%%%%%%%%%%%%%%%%%%%%
%%%%%%%%%%%%%%%%%%%%%%%%%%%%%%%%%%%%%%%%
\section{Radiative decays of the $a_{1}$ and $b_{1}$ mesons}

At first, we will consider the radiative 
decays of $a_{1}$ and $b_{1}$ mesons. 
In this work, we focus on
 the radiative transitions among the GS mesons. 
Therefore we are able to adopt simply the effective spin-type 
interaction, 
\beq
H= \ \bar{q} \
\sigma_{\mu\nu}F_{\mu\nu}((iv\gamma)g + g') \ q \ .
\label{em}
\eeq
Here we introduced two independent coupling 
parameters $g$ and $g^{'}$. 
The $g$ term contributes to only quark chirality conserving 
transitions, while  the $g^{'}$ term does to 
chirality non-conserving ones. 
By applying the quark-photon interaction ({\ref{em}}), 
the effective meson current is given 
by the following formulas, 
\beq
J_{\mu}^{}(P,P^{'})=J_{1,\mu}^{}(P,P^{'}) 
+J_{2,\mu}^{}(P,P^{'}).
\eeq
Here, subscript $1$ ($2$) represents the coupling of the 
emitted single photon with the relevant meson system 
through constituent quark (anti-quark). 
The specific form of the current is represented by 
\beq
J_{1,\mu}^{}(P,P^{'})&=&e_{q} I_{G}^{(\gamma)}~\langle 
\bar{W}^{(-)}(v^{'}) [2 g
i\sigma_{\mu\nu}q_{\nu}] 
iv\gamma W^{(+)}(v) iv\gamma\rangle
 \ ,
\\
J_{2,\mu}^{}(P,P^{'})&=&e_{\bar{q}} I_{G}^{(\gamma)}~\langle 
iv\gamma W^{(+)}(v){iv\gamma}[-2g^{}(-i\sigma_{\mu\nu}q_{\nu})]
\bar{W}^{(-)}(v^{'})\rangle
 \, 
\eeq
for the case of the chirality conserving transition; 
and similarly 
\beq
J_{1,\mu}^{'}(P,P^{'})&=&e_{q} I_{G}^{(\gamma)}~\langle 
\bar{W}^{(-)}(v^{'}) [2 g_{}^{'}
i\sigma_{\mu\nu}q_{\nu}]W^{(+)}(v)iv\gamma\rangle
 \ , \\
J_{2,\mu}^{'}(P,P^{'})&=&e_{\bar{q}} I_{G}^{(\gamma)}~\langle 
iv\gamma W^{(+)}(v)[-2 g_{}^{'}(-i\sigma_{\mu\nu}q_{\nu})]
\bar{W}^{(-)}(v^{'})\rangle
 \ ,
\eeq
for the case of the chirality 
non-conserving transition. 
Here $q_{\mu}=P_{\mu}-P^{'}_{\mu}$ denotes 
the 4-momentum of emitted photon, 
$I_{G}^{(\gamma)}$ is overlapping integral (OI) of space-time oscillator 
function, which gives a Lorentz invariant 
transition form factor as 
\beq
I_{G}^{(\gamma)}&=&\int d^4 x f_{G}^{*}(v^{'}, x) f_{G}(v,x) e^{-i\frac{1}{2}q_{\mu}x_{\mu}}\\
&=&(\frac{2MM^{'}}{M^2+M^{'2}}){\rm exp}[-\frac{1}{2\Omega}
\frac{(M^2-M^{'2})^2}{M^{2}+M^{'2}}], 
\label{IGgamma}
\eeq
where we introduce the parameter $\Omega$ corresponding to the 
Regge slope inverse. \\
In our scheme the relativistic covariance 
of the spin current, due to the inclusion 
of Dirac spinor with 
negative $\rho_{3}$-value, 
plays an important role in some radiative 
transition processes. 
To clarify this point, we rewrite the spin 
current vertex operator as 
\beq
\sigma_{\mu\nu} iq_{\nu}  A_{\mu}=\sigma_{\mu\nu} F_{\mu\nu} =
{\mbox{\boldmath{$\sigma$}}} \cdot {\bf{B}} - i\rho_{1} 
{\mbox{\boldmath$\sigma$}}\cdot{\bf{E}}~~.
%\sigma_{4j} iq_{j}  A_{4}=  
\eeq
In the cases of transition between both positive 
(negative) $\rho_{3}$ Dirac spinors, 
as is well known, the main contribution comes from 
the magnetic interaction. 
On the other hand, in the case of transitions between Dirac spinors 
with positive and negative $\rho_{3}$-values, 
the electric interaction, coming from the 
$\sigma_{i4}iq_{i} A_{4}$-term, becomes a dominant contribution. 
As a results, this {\it intrinsic electric dipole}{\cite{Dshad03}} 
transition gives an important role 
for the transition accompanied with their parity change, 
such as $a_{1} (b_{1}) \to \pi \gamma$ processes. 
\\
In this work, we take the following values of parameters in our scheme.
\begin{itemize}
\item ($g$, $g_{}^{'}$)=($2.59$, $1.40$) \ from \ $\Gamma_{\rm EXP}(
b_{1}^{+} \to \pi^{+} \gamma)$ and 
$\Gamma_{\rm EXP}(\rho^{+} \to \pi^{+} \gamma)$ 
\item $\Omega_{n\bar{n}}
=1.13 \ {\rm GeV}^2$ from 
$\Omega=M({}^{3}P_{2})^2-M({}^{3}S_{1})^2
=M(a_{2}(1320))^2-M(\rho (770))^2$
\end{itemize}
The masses of the respective mesons 
are taken from PDG{\cite{PDG2007}}, 
except for the one of the pion 
in the form factor with $M_{\pi}=0.78~{\rm GeV}$. 
The estimated widths are in comparison with experiment
in Table {\ref{tab3}}.
Results for this calculation are consistent with experiments. 
%%%%%%%%%%%%%%%%%%%%%%%%%%%%%%%%%%%%%%%
\begin{table}[t]
\centering
\caption{ {\it Radiative decay 
widths (KeV) in comparison with experiment.
Experimental data 
are taken from PDG{\cite{PDG2007}}.} 
}
\vskip 0.1 in
\begin{tabular}{|l|c|c|} \hline
Process &  Our results & Experimental values \\
\hline
\hline
 $\rho(770)\to\pi\gamma$   & 68~(input)  & 68$\pm$7 \\ \hline
 $b_{1}(1235)\to\pi\gamma$   & 230~(input)  & 230$\pm$60 \\ \hline
 $a_{1}(1260)\to\pi\gamma$   & 604  & 640$\pm$246 \\
\hline
\end{tabular}
\label{tab3}
\end{table}
%%%%%%%%%%%%%%%%%%%%%%%%%%%%%%%%%%%%%%%
\section{Pion emissions of $a_{1}$ and $b_{1}$ mesons}
Next we consider the strong decays with one pion emission. 
We adopt simply the 
following two types of effective quark-pion interactions; 
\beq
L_{ps}&=&g_{ps} \ \bar{q}(-i\gamma_{5})q 
\ \pi, \\
L_{pv}&=& g_{pv} \ \bar{q}(-i\gamma_{5}\gamma_{\mu})q 
\ \partial_{\mu} \pi.
\eeq
Note that here, $\pi$ ( and $\sigma$ ) meson is treated 
as an external local-field. 
Resultant matrix elements are given as a sum of two terms; 
\beq
T &=& T_{ps} + T_{pv} \ , \\
T_{ps} &=& g_{ps} I_{G}^{(\pi)}~\la W (v^{'}) (- i\gamma_{5} \pi) W (v) iv\gamma \ra  +
 c.c. \ \ , \\
T_{pv} &=& g_{pv} I_{G}^{(\pi)}~\la W (v^{'}) (-\gamma_{5} \gamma_{\mu}q_{\mu}\pi) 
W (v) iv\gamma \ra + c.c. \ \ . 
\eeq
In the above case, the OI of the space-time WF is given by 
\beq
I_{G}^{(\pi)}&=&\int d^4 x f_{G}^{*}(v^{'}, x) f_{G}(v,x) e^{-i\frac{1}{2}q_{\mu}x_{\mu}}\\
&=&(\frac{2MM^{'}}{M^2+M^{'2}-m_{\pi}^2})
{\rm exp}[-
\frac{(M^2-M^{'2})^2-m_{\pi}^2(M^2 + M^{'2})}
{2\Omega \left(M^{2}+M^{'2}-m_{\pi}^2\right)}], 
\eeq
where $q^2=-m_{\pi}^2$, $q_{\mu}=P_{\mu}-P^{'}_{\mu}$ 
being the 4-momentum of emitted pion. 
The relevant decay amplitude is 
\beq
T=f_{1}\ \epsilon_{\mu}(v')\epsilon_{\mu}(v) \ + \ 
f_{2} \ (q_{\mu}\epsilon_{\mu}(v'))(q_{\nu}\epsilon_{\nu}(v)).
\label{ampT}
\eeq
The explicit forms of $f_{1}$ and $f_{2}$ are shown in Table {\ref{tab4}}. 
It may be worthwhile to note that at least 
two coupling types ( expressed $f_{1}$ and $f_{2}$ in the above ) 
are required to reproduce the experimental data on $D/S$-wave amplitude ratios.\\
Our decay interaction contains two 
independent coupling parameters, $g_{ps}$ and $g_{pv}$, 
which will be commonly applied to all quark-pion vertices
{\footnote{
As an example, it is applied to the study of 
`extra'- $\kappa$ meson{\cite{Yamada}}.
}}. 
These are determined from the experimental data of 
$D/S$-wave amplitude ratio and total width of $b_{1}$ meson as, 
\begin{itemize}
\item $\frac{g_{ps}}{g_{pv}}=0.149 \ {\rm GeV}$ \ from
\ $T_{D}/T_{S}|_{\rm EXP}(b_{1}^{+} 
\to \omega \pi^{+}) = + 0.277$
\item $g_{pv}=14.0$ \ 
from \ $\Gamma_{\rm EXP}(
b_{1}^{+} \to \omega \pi^{+}) \approx 
\Gamma_{\rm EXP}(
b_{1}^{+}{}_{{\rm total}})=142 \ {\rm MeV}$. 
\end{itemize}
The masses of the relevant mesons 
are taken from PDG{\cite{PDG2007}}. 
The numerical results are shown in Table {\ref{tab5}}.~
%%%%%%%%%%%%%%%%%%%%%%%%%%%%%%%%%%%%%%%%
\begin{table}[htbp]
\centering
\caption{ \it 
Coefficients of decay amplitude ({\ref{ampT}}) for $a_{1}\to\rho\pi$ and 
$b_{1}\to\omega\pi$ process.
}
\vskip 0.1 in
\begin{tabular}{|l|c|c|} \hline
          &  $b_{1}\to \omega \pi $ & $a_{1}\to \rho \pi$ \\
\hline
\hline
 $f_{1}$   & $I_{G}\times
( -g_{ps}+(\omega M- M^{'})g_{pv})$ & $I_{G}\times
( -g_{ps}\omega+(M-\omega M^{'})g_{pv})$\\
 $f_{2}$   &$I_{G}\times
( -g_{pv}\frac{1}{M^{'}})$  &$I_{G}\times
( g_{ps}\frac{1}{M M^{'}}+g_{pv}\frac{1}{M})$ \\
\hline
\end{tabular}
\label{tab4}
\end{table}
%%%%%%%%%%%%%%%%%%%%%%%%%%%%%%%%%%%%%%%%
\begin{table}[htbp]
\centering
\caption{ \it Numerical results for pion emissions 
of $a_{1}$ and $b_{1}$ mesons. Experimental data 
are taken from PDG{\cite{PDG2007}}.
}
\vskip 0.1 in
\begin{tabular}{|l|c|c|c|c|} \hline
          &  \multicolumn{2}{|c|}{$T_{D}/T_{S}$} & \multicolumn{2}{|c|}{Width (MeV)} \\
\cline{2-3}\cline{4-5}
process &Our results&Experimental values&$\Gamma_{\rm partial}^{\rm theor.}$&
$\Gamma_{\rm total}^{\rm Exp.}$\\
\hline
\hline
 $b_{1}\to \omega \pi $   &0.277(input)  &
 0.277$\pm$ 0.027  &142(input)&142$\pm$9\\
 $a_{1}\to \rho \pi$  &-0.344 &
 -0.108$\pm$ 0.016&  191&
$250 \sim 600 $\\
\hline
\end{tabular}
\label{tab5}
\end{table}
%%%%%%%%%%%%%%%%%%%%%%%%%%%%%%%%%%%%%%%%
\section{Concluding remarks}
In this work, we investigate the decay properties of 
$q\bar{q}$ $S$-wave $a_{1}$ and $b_{1}$ mesons 
in the $\tilde{U}(12)$-scheme, 
by assigning them with $a_{1}(1260)$ 
and $b_{1}(1235)$ mesons, respectively. \\
At first, it is shown that the radiative decay widths of 
$(a_{1},b_{1},\rho) \to \pi \gamma$ processes are 
consistently reproduced by using the simple spin-type 
quark-photon effective interaction 
%with two coupling parameters 
in the framework of the $\widetilde{U}(12)$-scheme.\\
Secondly, for the strong one-pion emission decays, 
assuming the $ps$- and $pv$-type quark-pion effective 
interactions, the $D/S$-wave amplitude ratios and partial widths of 
$a_{1} (b_{1}) \to \rho (\omega)\pi $ decays 
are evaluated. 
As a results, by inputting the data for the $b_{1}$ meson, 
the sign of $D/S$-wave amplitude ratio for the 
$a_{1} \to \rho \pi $ decay agrees with the 
experiments, but its absolute value is about 
three time larger than experiment. 
Partial width of $a_{1} \to \rho \pi $ is predicted 
with $\Gamma ( a_{1} \to \rho \pi)\sim 200 {\rm MeV} $.\\
The interaction adopted in this work 
for the radiative/strong decays should be 
tested by applying it to other various decay processes. 
%%%%%%%%%%%%%%%%%%%%%%%%%%%%%%%%%%%%%%%%

%
\end{document}